\documentclass[nonacm, sigconf]{acmart}

\setcopyright{none}
\renewcommand\footnotetextcopyrightpermission[1]{}

\usepackage{enumitem}
\usepackage{pifont}
\newcommand{\cmark}{\ding{51}}%
\newcommand{\xmark}{\textcolor{red}{\ding{55}}}%

\usepackage{fancyhdr}
\fancypagestyle{plain}{%
   \fancyhf{} %
   \fancyfoot[L]{ACM SIGCOMM Computer Communication Review}%
   \fancyfoot[R]{Volume 48 Issue 1, January 2018}%
}
\fancypagestyle{firstpagestyle}{%
   \fancyhf{} %
   \fancyfoot[L]{ACM SIGCOMM Computer Communication Review}%
   \fancyfoot[R]{Volume 48 Issue 1, January 2018}%
}

\usepackage{balance}

\begin{document}

\title{SecureABC: Secure AntiBody Certificates for COVID-19}

\author{Chris Hicks*}
\affiliation{
	\institution{The Alan Turing Institute}
}
\email{chicks@turing.ac.uk}

\author{David Butler*}
\affiliation{
	\institution{The Alan Turing Institute}
}
\email{dbutler@turing.ac.uk}

\author{Carsten Maple}
\affiliation{
	\institution{The University of Warwick}
}
\email{cm@warwick.ac.uk}

\author{Jon Crowcroft}
\affiliation{
	\institution{The University of Cambridge}
}
\email{jon.crowcroft@cl.cam.ac.uk}

\begin{abstract}
COVID-19 has resulted in unprecedented social distancing policies being enforced worldwide. As governments seek to restore their economies, open workplaces and permit travel there is a demand for technologies that may alleviate the requirement for social distancing whilst also protecting healthcare services. In this work we explore the controversial technique of so-called immunity passports and present SecureABC: a decentralised, privacy-preserving protocol for issuing and verifying antibody certificates. We consider the implications of antibody certificate systems, develop a set of risk-minimising principles and a security framework for their evaluation, and show that these may be satisfied in practice. Finally, we also develop two additional protocols that minimise individual discrimination but which still allow for collective transmission risk to be estimated. We use these two protocols to illustrate the utility-privacy trade-offs of antibody certificates and their alternatives.

\end{abstract}



\maketitle
	
\section{Introduction}\label{sec:intro}

Governments worldwide are currently dealing with the Coronavirus pandemic, an outbreak of the SARS-CoV-2 virus which has already resulted in hundreds of thousands of deaths \cite{covid_tracker} and has caused unprecedented social changes for billions of people worldwide. Most countries have responded to the pandemic with policies aimed at enforcing social distancing, a technique certain to suppress the transmission of the virus when it is applied uniformly to the whole population \cite{Imperial_Report_9}. Whilst effective, social distancing measures come at a significant social and economic cost and may not be a feasible long-term policy option in many countries. In addition, it has been noted that the approach may have a disproportionate impact on disadvantaged groups in society \cite{unequal}. As the effective reproduction rate ($R_t$) is reduced below 1.0, and the number of COVID-19 cases falls, there are clear benefits to governments seeking technologies that may alleviate the requirement for population-wide social distancing whilst also protecting healthcare services and maintaining human and civil rights. In particular, measures such as intensive testing, contact tracing and selective isolation have been proposed \cite{trace_test_quarantine}.

A promising, but controversial \cite{Kofler_Baylis_2020}, technique for relaxing the need for population wide social distancing is the use of so called ``immunity passports'' or ``risk-free certificates''. In this work we adopt for the term ``antibody certificates'' which we believe to be more accurate. The general idea is that a test for antibodies to SARS-CoV-2, the virus that causes COVID-19, could serve as the basis for a certificate that frees an individual from the most restrictive social distancing regulations. Assuming that strong correlates of protection to SARS-CoV-2 can be identified, antibody certificates could provide a key technology in enabling the transition away from total lockdown. Chile and Estonia have already announced plans to issue and trial such certificatees, respectively, and policy makers in the USA, UK, Italy and Germany are also considering the approach \cite{chile_passport, uk_germany_passport}. In light of the interest already shown by several governments, and the emergence of a number of commercial solutions, an academic consideration of antibody certificates is called for. In particular it is important to provide detailed technical solutions to the problem, and to identity current limitations, so that policy makers can be properly informed. We preface our technical work with four risk-minimising principles which aim to address some of the concerns surrounding antibody certificates.




\subsection*{Contributions}
\begin{enumerate}
    \item We provide a framework of security and privacy requirements for antibody certificate systems. Our framework provides the basis for ensuring such systems have the correct properties and can be evaluated on common terms.
    
    \item We present our SecureABC (Secure AntiBody Certificate) protocol, a decentralised and privacy-preserving solution for antibody certificates. We evaluate SecureABC with respect to our framework of security and privacy requirements. More generally, SecureABC is applicable to the emerging field of health certification and provides a secure mechanism for certifying health data.
    
    \item We provide evidence that SecureABC is efficient and practical with our proof of concept implementation which includes an Android application for verifying SecureABC QR codes.
    
    \item Finally, we explore the possibilities and limitations of technically mitigating individual health discrimination with two protocols that do not depend on good policy. We evaluate the trade-offs of these two protocols, and our SecureABC solution, with respect to their utility and privacy.

\end{enumerate}

\subsection*{Outline}

First in Section \ref{sec:general_principles} we review the concerns surrounding antibody certificates and derive four general principles that may help to minimise the risk posed by these systems. Next, in Section \ref{sec:model_and_prelims} we define our protocol model and introduce the required cryptographic primitives. Section \ref{sec:secureabc} presents the full details of our SecureABC solution. Then, in Section \ref{sec:properties}, we define our framework of security properties for antibody certificates and evaluate SecureABC with respect to them. Section \ref{sec:implementation_and_performance} shows the results of our implementation of SecureABC. Section \ref{sec:non-discrimination} details two protocols that protect individuals from immunity-based discrimination. Finally, Sections \ref{sec:related_work} and \ref{sec:conclusion} review related work and conclude respectively. 

\section{Antibody Certificate Principles} \label{sec:general_principles}


In this section we build towards a set of general principles for minimising the risk of an antibody certificate system. Evidence is mounting that many nations \cite{chile_passport, uk_germany_passport} are at least considering COVID-19 antibody certificates, so it is important that techniques for minimising the risks posed by such systems are developed. To aid and direct our discussion we begin with a general use case in which citizens wishing to travel on public transport are required to present a certificate affirming that they have some level of protective immunity to the SARS-CoV-2 virus.

\vspace{1ex}
\noindent\textbf{Use case 1: basic use case.} A healthcare provider will test citizens for antibodies to COVID-19. If antibodies are found, then an antibody certificate will be issued. The citizen can now present their certificate to the transport service provider who, after establishing the authenticity of the certificate, will allow access to transportation.

\subsection{Antibody Certificate Concerns}

Antibody certificates have attracted significant criticism. Here we condense the main concerns of Kofler et al. \cite{Kofler_Baylis_2020} and the WHO \cite{WHO_immunit_passports}:

\begin{description}[leftmargin=0ex, itemsep=1.5ex]
    \item[Alterations to scientific advice.] As a novel disease, COVID-19 is not yet well understood by the scientific community \cite{Lewis_2020}. For example, the WHO express concern that the presence of antibodies is not an accurate indicator of immunity and state that such tests ``need further validation to determine their accuracy and reliability'' \cite{WHO_immunit_passports}.
    
    \item[Discrimination.] Antibody certificates will create an attribute for discrimination \cite{uk_germany_passport} by establishing a clear distinction between those with antibodies and those without. A report by the Ada Lovelace Institute emphasises that ``Discrimination and stigmatisation may become commonplace if immunity becomes an element of identity'' \cite{lovelace_report}. Additional inequality could be introduced if tests are not universally and freely available to all -- antibody certificates, and the associated freedoms, could become a luxury of the rich.
    
    \item[May negatively impact behavior.] An antibody certificate could become synonymous with freedom as those in possession could be allowed access to ``the post-lockdown world'' while those without could remain subject to social distancing policies. This may create a strong incentive for people to attempt to obtain a certificate by any means. Coupled with the belief that having, and surviving, the virus corresponds to gaining immunity \cite{sars_lessons, reinfects}, it is hypothesised that antibody certificates may actively encourage people to try to become infected. It has also been shown that believing one has immunity could lead people to behave less cautiously \cite{ Wallere040448}.
    
    \item[Feature creep.] In response to COVID-19, Governments are exercising a heightened level of control and surveillance over their populations \cite{democracy, surveillance2}. Antibody certificate systems could provide authorities with an opportunity to implement technologies, and to collect data, that could have unanticipated negative consequences on society \cite{lovelace_report}.
\end{description}

These concerns are persuasive in advocating against antibody certificates in the basic use case setting we have described. We are therefore motivated to address these concerns with a set of general principles which aim to minimise, where possible, the risks associated with antibody certificates.

\subsection{Risk Mitigating Principles} 
Principles for minimising the risk of antibody certificates.

\begin{description}[leftmargin=0ex, itemsep=1.5ex]
    \item[Rename, educate and allow revocation.] Until there is strong evidence of protective immunity (and its correlates) to COVID-19, the word ``immunity'' should not be used. This is misleading at best and dangerously inaccurate at worst -- creating a false sense of security \cite{Wallere040448}. We opt for ``antibody certificate'' as it reflects the function more accurately. We note that in \cite{eithics_liicences} Persad and Emanuel also advocate for a name other than ``immunity passports'', although we would argue that their ``immunity-license'' is still somewhat misleading. Beyond semantics, appropriate levels of public education with regard the benefits and limitations of antibody certificates is vital. Indeed, the Information Commissioners Office (ICO) strongly advocate being transparent about the purpose and benefit of the closely related contact tracing technology \cite{ico_report_on_contact_tracing}. Finally, we call for systems which support efficient and fast revocation of certificates and service providers. This is essential to maintain pace with the fast-changing scientific understanding and dynamic policies which relate to COVID-19.
    
    \item[Access to testing and technology.] Wealth, location and demographic profile must not impact access to obtaining or using a certificate. In other words, beyond their specified purpose, antibody certificates must avoid any additional discrimination. The Ada Lovelace Institute report \cite{lovelace_report} similarly calls for ``measures for ensuring vulnerable groups are not excluded from the operation of the system'' in relation to antibody certificates. In the case of contact tracing, the ICO state that ``special consideration for different societal groups'' is paramount.
    
    \item[Proportional and necessary use.] As an inherently discriminatory technology we recognise that antibody certificates must be used cautiously. We advocate that restrictions imposed by antibody certification upon freedom of movement, or access to services, must be strictly necessary and proportional to the specific risks and harms of COVID-19 transmission in each particular use-case. We note that in their draft Coronavirus Bill [44] Edwards et al. emphasise that proportionality is the key to reducing discrimination. To ensure that verifiers behave according to agreed guidelines, mechanisms should be in place which allow a user to determine that a service provider is authorised before presenting their certificate. Such ``mutual authorisation'' between users and service providers will provide a higher degree of confidence to users.
    
    \item[Maintain user ownership of data.] An antibody certificate system should be designed in such a way that users can, at all times, control the use of their data. In particular, users must control when and where to use their data to demonstrate their test result. This is aligned with the ICO's principle of ``giving users control'' \cite{ico_report_on_contact_tracing} with regard to contact tracing.
\end{description}


These principles, in particular ensuring the necessity and proportionality of antibody certificates for each use case, likely rule out our basic use case. We hence provide two additional use cases that may present a more favourable balance between the risks of antibody certificates and the benefit of their application.


\vspace{1ex}
\noindent\textbf{Use case 2: indicating user risk.} Consider the basic use case where users of a public transport system are required to show their antibody certificate before making a journey. Under our risk minimising principles we advocate that certificates are still checked; however, access is granted to all. In this scenario, antibody certificates can still provide a benefit by allowing testing, cleaning and other mitigating techniques to be optimised based on the relative use of each service by untested and tested-negative citizens. 

\vspace{1ex}
\noindent\textbf{Use case 3: helping vulnerable members of society.}
When providing food delivery services for vulnerable people who cannot safely leave their home, preferring the carrier to hold an antibody certificate could reduce the risk faced by the recipient. This scenario, in which the carrier would hold an antibody certificate and her employer verify that it is valid, could represent a necessary and proportional trade-off between restricting access to work and minimising the risk faced by vulnerable groups.

\section{Model and Preliminaries} \label{sec:model_and_prelims}

\begin{figure}[h]
    \centering
    \includegraphics[scale=0.21]{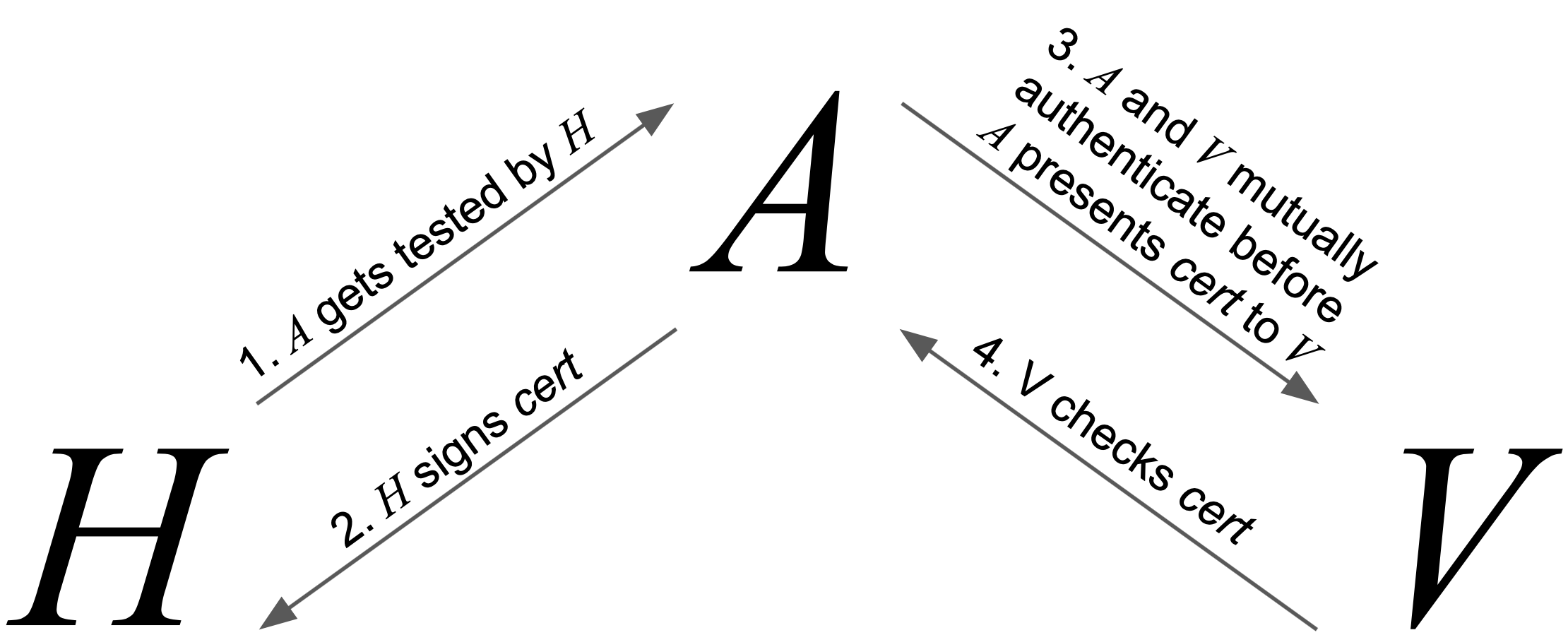}
    \caption{Our general model of an antibody certificate protocol run between a healthcare provider, a user and a service provider.}
    \label{fig:system_outline}
\end{figure}

Our antibody certificate protocol model comprises three parties. We denote the Healthcare provider as Harry ($H$), the citizen or user of the scheme as Alice ($A$) and the service provider, who Verifies user certificates, as Verity ($V$). Our general model is outlined in Figure \ref{fig:system_outline}: in step (1) $A$ gets tested for antibodies by $H$. In step (2), if the test detects an appropriate number of antibodies, $H$ signs an antibody certificate which is sent to $A$. In step (3), after mutually authenticating, $A$ presents her certificate to $V$ and then finally, in step (4), $V$ checks the authenticity (i.e. that is was created by $H$) and attributes of the certificate (e.g. by matching a photo and name to $A$). Our system model assumes a root of trust in the system, for example the government, that implicitly authorises both $H$ and $V$ as legitimate entities. A root of trust is needed because a verifier $V$ cannot reasonably be expected to know in advance all of the healthcare providers that can be trusted to authorise antibody certificates. Similarly, the user $A$ cannot be expected to always know which particular verifiers can be trusted to verify their antibody certificates. 

\subsection{Preliminaries}

SecureABC requires both a secure public key signature scheme and a secure public key encryption scheme. Both of these schemes are made up of three algorithms. The first algorithm generates a public key pair, the second either signs or encrypts and the third verifies the signature or decrypts the ciphertext, respectively. We denote the signature scheme by the tuple $(\mathit{key\text{-}gen_{sign}},  \mathit{sign}, \mathit{verify})$ and the encryption scheme by the tuple $(\mathit{key\text{-}gen_{enc}},  \mathit{enc}, \mathit{dec})$.

We use a signature scheme to allow the healthcare provider to sign antibody certificates and therefore, intuitively, we require the signature scheme to be unforgeable. Specifically, we require it to be EUF-CMA (Existential Unforgeability under Chosen Message Attack) secure. Rather than define this standard property here, we refer the reader to Goldwasser et al. \cite{DBLP:journals/siamcomp/GoldwasserMR88} for the common definition. We note that the standard Elliptic Curve Digital Signature Algorithm (ECDSA) \cite{FIPS186} is sufficient for our needs.

We use the encryption scheme in the authentication phase for app-based users. Here the user encrypts their signed certificate such that only an authorized, non-revoked service provider can decrypt it (we discuss this property in more detail when evaluating our scheme in Section \ref{sec:properties}). Our particular use of an encryption scheme means that any public key encryption scheme that is at least IND-CPA secure will suffice. For a standard definition of IND-CPA we refer the reader to Katz and Lindell \cite{ind-cpa}.

\section{S\texorpdfstring{\MakeLowercase{ecure}}AABC} \label{sec:secureabc}

In this section we present SecureABC, our secure antibody certificate scheme that realises the model defined in Section \ref{sec:model_and_prelims}. SecureABC is a distributed, privacy-preserving antibody certificate protocol that allows for both paper-based and app-based user credentials. Providing printable, paper-based credentials is important because even in the most developed countries, the adoption of smartphones is not absolute. Indeed, requiring the use of electronic devices may exclude vulnerable user groups \cite{unequal} and limit the reach of any deployment. We seek to provide strong privacy guarantees regardless of whether a user has a device capable of displaying a digital passport or not. In particular we seek to replicate the privacy of traditional identity documents, such as driver's licenses, which do not notify the issuer each time they are presented.

As noted in Section \ref{sec:model_and_prelims} our model assumes a common trust anchor, such as the government, that authorises both $H$ and $V$ as legitimate entities. In practice this means that every time $H$ or $V$ generate a new public key, the government signs it to indicate that it can be trusted. For brevity we implicitly assume this to be the case for all relevant keys.

\subsection*{High level overview of SecureABC}\hspace{1ex}\\
The SecureABC protocol comprises three phases: Setup, Issue and Authentication. The Setup protocol is run by $H$, who generates the public key pair for the signature scheme he will use to sign antibody certificates. Next, the Issue phase is run between $H$ and $A$. At the end of the Issue phase $A$ receives an antibody certificate, signed by $H$, which she can use to demonstrate her antibody status to service providers. Finally, the Authentication phase is run between $A$ and $V$. This phase allows $V$ to convince $A$ that she is an authorised service provider and allows $A$ to convince $V$ that she has a valid antibody certificate. We provide two Authentication phase sub-protocols which allow $A$ to choose between using either a paper-based or an app-based credential.

\subsection*{The SecureABC Protocol}\hspace{1ex}\\
The Setup, Issue and Authentication phases of the SecureABC protocol are as follows.

\vspace{1ex}
\noindent \textbf{Setup}

\noindent $H$ initialises the set of revoked Certificate ID ($(\mathit{CID}$) numbers $\mathit{rev} = \{\}$, the set of revoked verifier public keys $\mathit{rev_{V}} = \{\}$ and generates the public and private key for the signature scheme: $$(pk_H, sk_H) \leftarrow \mathit{key\text{-}gen_{sign}}(\cdot)$$

\vspace{1ex}
\noindent \textbf{Issue}

\noindent $H$ interacts with $A$ to issue a signed antibody certificate.

\begin{enumerate}
    \item $A$ is tested for antibodies by $H$, who records the test identity number ($\mathit{TID}$) that produced the test result. If the test is positive for antibodies, and $A$ has not already been issued a certificate, then $H$ generates a random $(\mathit{CID}$, initialises a corresponding revocation bit $b_{\mathit{CID}} = \mathit{False}$ and specifies a validity period $\mathit{date}_{\mathit{CID}}$. $H$ stores $(\mathit{CID}, \mathit{TID}, b_{\mathit{CID}},\mathit{date}_{\mathit{CID}})$ informs $A$ of her test result.
    
    \item $A$ provides $H$ with a photograph, $\mathit{photo_A}$, and a communication channel, $\mathit{comm_A}$, which will be used to send passport status updates (e.g. upon revocation or test recall). In practice, $A$ will choose one of a small number of communication options such as SMS, email or post.
    
    \item $H$ appends $\mathit{comm_A}$ to the record he holds for $A$ and then sends her the signed antibody certificate $\mathit{cert_A}$ which is computed as follows:
    $$\mathit{cert_A} = \mathit{sign}_{pk_{H}}(\mathit{name_A}, \mathit{photo_A},  \mathit{date_{CID}}, \mathit{CID})$$
\end{enumerate}

Certificates can either be paper-based or app-based, in Section \ref{sec:implementation_and_performance} we show how QR codes can be used to encode such certificates. In the remainder of this paper we assume this encoding.

In the interests of mitigating economic discrimination, we recommend that if a user is unable, or unwilling, to provide a photograph then they are able to obtain one of the required standard for free during the issue phase. One option could be to create a mobile app that can be used by $H$ to take such photographs on behalf of each user, it is important however that any such image is deleted by $H$ as soon as the certificate has been issued. All of this could be handled automatically by a suitable audited application.

\vspace{1ex}
\noindent \textbf{Authentication}

\noindent $A$ interacts with $V$ to demonstrate the authenticity and ownership of her antibody certificate. Authentication must be mutual, that is first $A$ must be convinced that $V$ is authorised by the government before allowing her to verify her certificate. $A$'s certificates are either paper-based or app-based we require two different authentication sub-protocols which correspond to the manual and automated authentication of $V$ by $A$, respectively. In both authentication sub-protocols, $V$ runs an app which she uses to scan and verify $A$'s credential. The app also periodically downloads the list of revoked $\mathit{CID}$ numbers $\mathit{rev}$ and $H$'s public key $pk_{H}$. The $\mathit{rev}$ list should be signed by $H$ to demonstrate its authenticity. We benchmark the performance of this app in Section \ref{sec:implementation_and_performance}.

\vspace{1ex}
\noindent \textbf{Paper-Based Authentication}

\noindent $A$ has $\mathit{cert_A}$ and $V$ has the list of revoked $\mathit{CID}$ numbers $\mathit{rev}$ and $H$'s public key $pk_H$.

\begin{enumerate}
    \item $A$ must manually convince herself that $V$ is a government-authorised service provider, for example based on context or viewing an identity document. If this step fails $A$ aborts the protocol.
    
    \item $A$ shows her certificate $\mathit{cert_A}$ to $V$.
    
    \item $V$ verifies $\mathit{cert_A}$. That is she confirms $\mathit{verify_{pk_{H}}}(\mathit{cert_A}) = \texttt{True}$, and learns $\mathit{name_A}$, $\mathit{photo_A}$, $\mathit{date_{CID}}$ and $\mathit{CID}$. She checks that $\mathit{CID}$ has not been revoked i.e. $\mathit{CID} \notin \mathit{rev}$, then compares $\mathit{photo_A}$ to $A$ and ensures that $\mathit{date_{CID}}$ has not elapsed. Optionally, if she deems it necessary for verification, $V$ may ask to see a second document bearing $\mathit{name_A}$ --- this process would happen offline\footnote{We envisage $V$ asking for this second factor only if she was unconvinced the photo provided strong binding, or if the use case demanded extra precautions to ensure the user was in fact the holder of the certificate.}.
\end{enumerate}

\vspace{1ex}
\noindent \textbf{App-Based Authentication} 

\noindent $A$ runs an app that stores her antibody certificate $\mathit{cert_A}$ and which periodically downloads a list of revoked verifier public keys, $rev_V$, from the government. $V$ has a list of revoked $\mathit{CID}$ numbers, $H$'s public key $pk_H$, an encryption public key pair $(pk_V, sk_V) \leftarrow \mathit{key\text{-}gen_{enc}}(\cdot)$ and a public-key certificate signed by H, $\mathit{cert}_V \leftarrow \mathit{sign}_{sk_H}(pk_V)$.
\begin{enumerate}
    \item $V$ sends $\mathit{cert}_V$ to $A$. In practice, $A$ uses the app to scan a QR code or read an NFC tag provided by $V$. $A$ verifies $cert_V$ and ensures is not on the revocation list, i.e. $pk_V \not\in \mathit{rev}_V \land \mathit{verify}_{pk_H}(cert_V)=\texttt{True}$. If either of these steps fail, then $A$ aborts the protocol.
    \item $A$ computes $\mathit{cert}^\prime_{A} = \mathit{enc}_{pk_V}(\mathit{cert}_A)$ which is displayed as a QR code and scanned by $V$.
    
    \item $V$ decrypts $A$'s certificate $\mathit{cert}_A = \mathit{dec}_{sk_V}(\mathit{cert}^\prime_A)$ and then proceeds as in Step 3 of the paper-based authentication protocol.
    
\end{enumerate}

If either the paper-based or app-based authentication protocol succeeds then $V$ accepts $A$'s antibody certificate, otherwise she does not. It is important to note that the use of encryption in our app-based authentication protocol is an opportunistic enhancement which provides slightly improved security properties for app-users. More details on this point are provided in our evaluation which follows. While not strictly part of the protocol we next detail how revocation would be achieved.

\vspace{1ex}
\noindent \textbf{Revocation.} 

\noindent $H$ periodically computes, signs and distributes $\mathit{rev}$ (e.g. daily) such that it comprises all $\mathit{CID}$ numbers in his private store that have the revocation bit $b_{\text{CID}} = \texttt{True}$. We claim this is sufficient for efficient revocation of certificates and, we illustrate the exact procedures for different use cases in our evaluation which follows.

\section{Security Properties and Evaluation} \label{sec:properties}

In this section we first present a general framework of security properties for the evaluation of antibody certificate protocols and then we then evaluate SecureABC with respect to these properties and also the general principles which we set out in Section \ref{sec:general_principles}.

\subsection{Security Properties} \label{sec:desired_properties}

Here we provide a basic set of properties for evaluating the security of antibody certificate systems. It is unlikely that any scheme can simultaneously satisfy all of these properties as several of them present a trade-off. For example, there is an inherent compromise between the anonymity of user certificates and the binding between the user and their certificate. We provide additional detail for some properties to help guide the reader.

\begin{description}[leftmargin=2ex, itemsep=1ex, topsep=2ex]
    \item[Forge and Tamper Proof.] \textit{We say the system is forge proof if a user cannot create a valid certificate alone. That is valid certificates can only be created for $A$ by $H$. Similarly, we say the system is tamper proof if the value of the attributes associated with a certificate cannot be altered by the user after the certificate has been issued.}
    
    \item[Binding.] \textit{We require that the only certificate a user can successfully use is the one that is assigned to them and which has not been revoked.}
    
    \item[Uniqueness.] \textit{Uniqueness is satisfied if a user can have at most one valid certificate associated to them at any one time.} -- We note that this property is important when binding is imperfect, for example when considering twins that may not share the same antibody test result.
    
    \item[Peer-Indistinguishability.] \textit{We define a peer as an unauthorised service provider (for example, a malicious citizen). We require that a peer cannot learn any information about the user from viewing the certificate.} -- Intuitively, peer-indistinguishability ensures that a user cannot be pressured into revealing their certificate by anyone except an authorised service provider, this alleviates the ``bully on the bus'' problem.
    
    \item[Unlinkability.] We consider unlinkability from the healthcare provider and the service provider in turn.
    \begin{itemize}[topsep=1ex]
        \item From healthcare provider: \textit{The healthcare provider cannot link a user to their authentication phase interactions.}
        \item From service provider: \textit{The service provider cannot link a user to a previous authentication phase interaction.} 
    \end{itemize}
    
    \item[Revocation of certificates.] \textit{We require that a user’s certificate can be revoked by the issuer.} Certificates may be invalidated for a number of reasons, we give three examples below:
    \begin{itemize}[topsep=1ex]
    \item Loss and theft: If a certificate becomes lost or compromised.
    \item Error: If a batch of tests are recalled because they were incorrect.
    \item Misuse: If evidence of certificate misuse is presented.
    \end{itemize}
    
    \item[Revocation of service providers.] \textit{We require that a service provider can be revoked from the list of authorised providers.} -- an authorised service provider may be revoked in the following situations:
    \begin{itemize}[topsep=1ex]
        \item Change of policy: A change in government policy may mean some service providers are no longer authorised.
        \item Sanctioning: If a service provider is deemed to not be following recommended guidelines it may lose its authorised status.
    \end{itemize}
    
\end{description}

\subsection{Evaluation}

Here we evaluate the SecureABC system in relation to the security properties and then with respect our general principles from Section \ref{sec:general_principles}. Table \ref{table:paper_vs_app_properties} compares the properties observed by the paper-based and app-based authentication mechanisms of SecureABC. In this work we only aim to provide an intuition for each property rather than a rigorous proof.

\begin{description}[leftmargin=2ex, itemsep=1ex, topsep=2ex]
    \item[Forge and Tamper Proof.] Here we consider the theoretical enforcement of the forge and tamper proof properties in SecureABC. Under the standard assumption that the underlying signature scheme is EUF-CMA secure, SecureABC is both forge and tamper proof. At a technical level the forge proof property is reduced to the unforgeability of the signature scheme. Moreover, the tamper proof property can also be reduced to the forge proof property as, in order to tamper with the value of the certificate, the user must also forge the corresponding signature.
    
    \item[Binding.] Alice is bound to her certificate by the photograph and name that are signed by Harry. This can be considered a strong binding property and is in line with current socially accepted verification mechanisms such as photocard driving licenses.
    
    \item[Uniqueness.] In the Issue phase Harry checks that to see if Alice currently has a valid certificate. Consequently, uniqueness is satisfied to the degree that it is already assured for medical record keeping.
    
    \item[Peer-Indistinguishability.] \hspace{1ex}
    
    \begin{itemize}[topsep=1ex]
        \item Paper-based: The paper-based authentication sub-protocol provides poor peer indistinguishability for users. In particular, the protocol only guards against non-technical peers without the ability to scan QR codes.
    
        \item App-based: The app-based authentication sub-protocol realises the peer indistinguishability property. This is achieved by enforcing mutual authentication between Alice and Verity. Here Verity is required to present a valid public key, signed by the government, which has not been revoked. Recall that since Verity's public key is signed using an EUF-CMA secure signature scheme, peer indistinguishability can be reduced to the difficulty of forging a government signature.
    \end{itemize}
    
    \item[Unlinkability.] \hspace{1ex}
    
    \begin{itemize}[topsep=1ex]
        \item From healthcare provider: SecureABC is decentralized after the issue phase. More precisely, the healthcare provider (or any central authority for that matter) is not involved in the protocol after the issue phase. Consequently to link the user to an authentication the healthcare provider would have to collude with the service providers, we assume this does not happen.
        \item From service provider: Since the service provider learns when a user authenticates with them, this property is not satisfied. If multiple service providers collude, then the colluding group all learn the linking.
    \end{itemize}

    \item[Revocation of Certificates.] Recall that Harry periodically computes, signs and distributes $\mathit{rev}$ (e.g. daily) such that it comprises all $\mathit{CID}$ numbers in his private store that have the revocation bit $b_{\mathit{CID}} = \texttt{True}$. Then, for the use-cases in Section \ref{sec:general_principles}, revocation of certificates can be realised as follows:
    \begin{itemize}[topsep=1ex]
        \item Loss: If Alice's certificate becomes lost or compromised, she must inform Harry. Harry looks up her $\mathit{CID}$ and sets the revocation bit $b_{\mathit{CID}} = \texttt{True}$ in his private store.
        
        \item Error: If a test result is recalled due to clinical error, Harry uses the $\mathit{TID}$ number to identify the corresponding $\mathit{CID}$ in his private store and then sets the revocation bit $b_{\text{CID}} = \texttt{True}$.
        
        \item Misuse: If evidence of certificate misuse3 emerges, a trusted authority (e.g. a court) should inform Harry of the $\mathit{CID}$ that was misused. Harry then sets the corresponding revocation bit $b_{\text{CID}} = \texttt{True}$.
    \end{itemize}
    
    \item[Revocation of Service Providers.] \hspace{1ex}
    \begin{itemize}[topsep=1ex]
    
    \item Paper-based: Our paper-based authentication does not provide technically enforced revocation of service providers. This property can only be obtained if a user is constantly educated as to which service providers are no longer authorized. For example, daily updates on coronavirus from the UK government have included details on which services and shops are allowed to operate. 
    \item App-based: This property is satisfied by the app-based authentication sub-protocol. Authentication only succeeds if, in the mutual authentication phase, the service provider sends the user a public key $pk_V$ that is signed by the government and which is not on the list of revoked verifiers $\mathit{rev_V}$. Since the user encrypts their antibody certificate using $pk_V$, the verifier must have the corresponding private key $pk_V$.
\end{itemize}
    
\end{description}

\begin{table}[h]
\centering
\begin{tabular}{l|c|c}
                        & Paper-Based & App-Based \\
\hline
Forge and Tamper proof       &       \cmark      &     \cmark    \\
Binding &       \cmark      &     \cmark    \\
Uniqueness       &       \cmark      &     \cmark    \\
Peer-Indistinguishability             &      \xmark      &     \cmark    \\
Unlinkability by $H$            &     \cmark        &     \cmark    \\
Unlinkability by $V$             &     \xmark        &     \xmark    \\
Revocation of Certificates &      \cmark       &      \cmark   \\
Revocation of Service Providers        &       \xmark      &       \cmark 
\end{tabular}
\vspace{2ex}
\caption{Security properties of the paper-based and app-based SecureABC antibody certificates.}
\label{table:paper_vs_app_properties}
\vspace{-3ex}
\end{table}

\subsection{SecureABC and the General Principles}

Here we position SecureABC with respect to the general principles we formulate in Section \ref{sec:general_principles}. Some of the components of our general principles, for example educating the public, are naturally out of scope when presenting a technical solution to antibody certificates; thus, we consider only those aspects that are relevant to the SecureABC protocol.

\begin{description}[leftmargin=2ex, itemsep=1ex, topsep=2ex]
\item[Rename, educate and allow revocation.] We follow our naming principle. SecureABC provides Secure AntiBody Certificates and implies neither immunity nor freedom of travel. Moreover SecureABC provides efficient and accurate revocation of both user certificates and, for app users only, service providers.

\item[Access to testing and technology.] We address digital exclusion in SecureABC by providing both a paper-based and app-based authentication protocol. In addition, we prescribe that Harry takes the photo of Alice in the issue phase. These design decisions ensure that access to (or willingness to use) technology including a ``passport photo machine'' does not discriminate against certain user groups.

\item[Proportional and necessary use.] Whilst this is ultimately a policy decision, we suggest two potentially proportional use cases for SecureABC in Section \ref{sec:general_principles}.

\item[Maintain user ownership of data.] After the initial issue phase, SecureABC is a decentralised system in which users authenticate directly with service providers. This keeps the user in control of when their data is used, and for what purpose, and may both minimise feature creep by governments and facilitate dismantling the system.
    
\end{description}

\section{Implementation and Performance} \label{sec:implementation_and_performance}

In this section we review the implementation and practical considerations of the SecureABC antibody certificate protocol and present the results of our reference implementation.

\subsection{QR Codes}
In our proof of concept implementation we have chosen to use QR codes to represent the signed antibody certificates that are issued to users. QR codes are a reasonable candidate for this purpose because they are a widely adopted, mature technology that offers both machine-readability and error tolerance. There are a range of libraries suitable for reading and writing QR codes, users understand how to interact with them, and they are resistant to wear and tear when printed. Attacks on QR code systems are surveyed by Krombholz et al. \cite{qr_code_survey}.

The storage capacity of a QR code is determined by a “symbol version” number between 1 and 40. Higher symbol numbers correspond to a higher storage capacity. The maximum storage capacity for a standard QR code is 2953 bytes \cite{qr_codes}. If a higher storage capacity is desired, for example for a higher resolution user photograph, our SecureABC protocol readily supports alternative technologies such as multi-layered QR codes \cite{6_layer_QR, colour_QR_codes} and Microsoft High Capacity Color Barcode (HCCB) \cite{hccb}.

\subsection{Proof of Concept} 

\begin{figure}[h]
    \centering
    \includegraphics[scale=0.29]{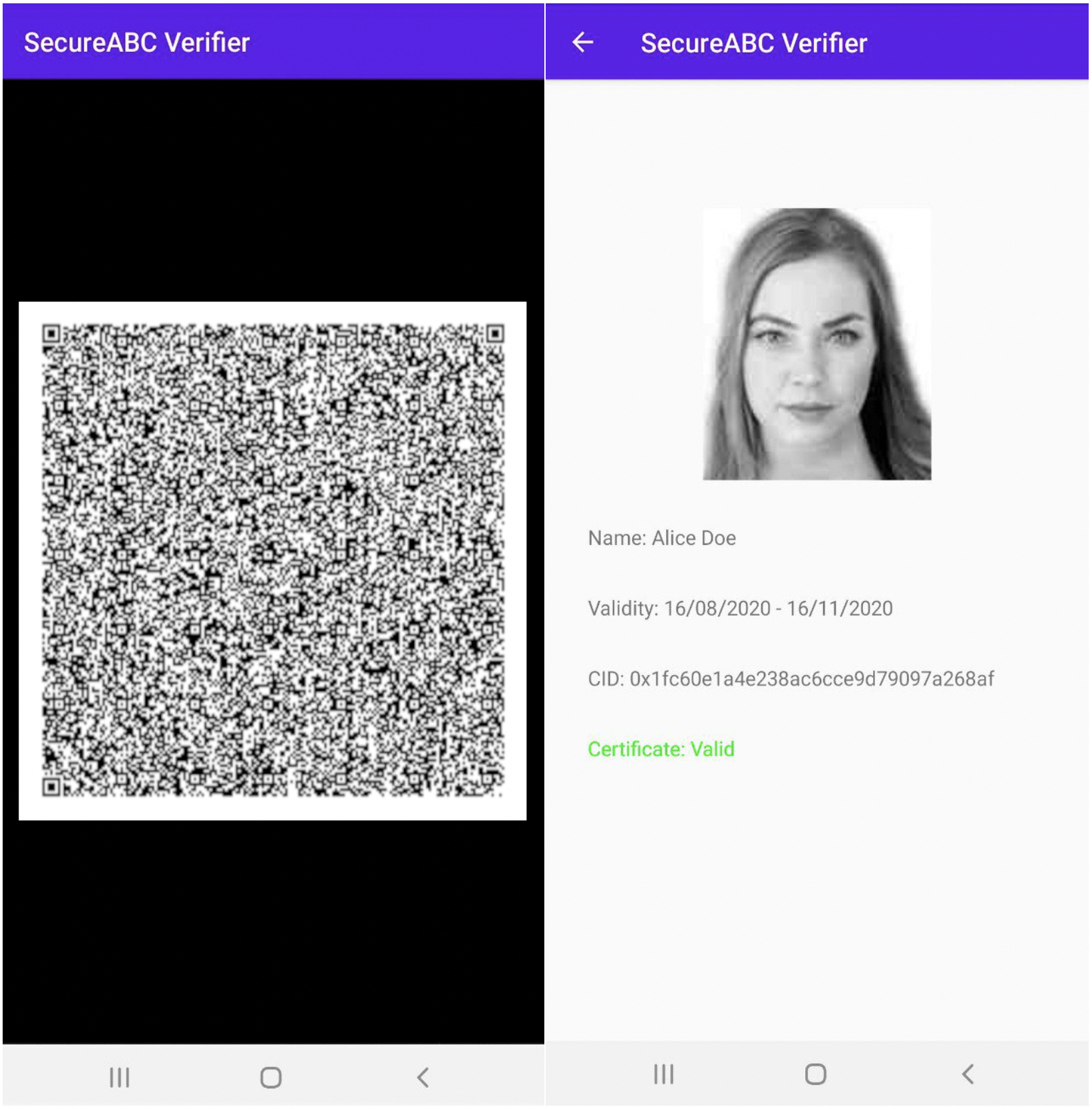}
    \caption{The SecureABC scanning screen (left) and verification screen (right) from our reference Android implementation. The scanning screen shows a QR code output during the issue phase of our protocol. As shown on the verification screen, this QR code comprises the user image \cite{nvidea_style_gan}, a name, a $\mathit{CID}$, a validity period and a 521-bit ECDSA signature.}
    \label{fig:user_qr}
\end{figure}

We have created an open source proof of concept implementation of the critical SecureABC components which can be found at \cite{SecureABC_github}. Firstly, our implementation allows for user credentials to be signed and compiled as a QR code as shown in Figure \ref{fig:user_qr}. This corresponds to the issue phase of our protocol. Secondly, we have implemented an Android application that runs the authentication phase of our protocol. In more detail, the application can read a SecureABC QR code, verify the cryptographic signature and display the user details as also shown in Figure \ref{fig:user_qr}. 

\subsection{Performance and Scalability} 

Here we first provide benchmarks for our implementation and then highlight some limiting factors and initial insights regarding optimisation. Finally, we consider the scalability and usability of our proposal.

SecureABC only requires highly efficient cryptographic algorithms; both signing and verification of antibody certificates are simply the standard ECDSA algorithms. Using our Python implementation on an Intel Core i9 Macbook, complete end-to-end SecureABC QR code generation takes less than 8 seconds. Recall that this is a one-time operation and, although this performance is already persuasive, we note that less than 2 seconds are spent actually signing the certificate data and the remainder is spent on the generation of the QR code. We discuss this further when considering optimisations below.

While issuance is a one-time process, user authentication occurs many times and is the more performance critical component. The results of our Android implementation show that a 521-bit ECDSA signature using the NIST standard curve P-521 can be verified in 3 ms using a Samsung Galaxy S9+ smartphone.

Next, we discuss our initial findings regarding optimisation. Our reference implementation allowed us to both better understand the potential of digital barcode technologies and recognise the specific limitations of QR codes. Firstly, encoding binary data into a QR code is beyond the specification of most standard libraries. This limitation required us to encode the user photograph using a printable representation before writing to a barcode. In particular our Android implementation requires base 64 encoding the user photograph before it is compiled into a QR code. This adds a very significant 33\% overhead to the number of bytes that presently need to be included in the QR code. While our current implementation is sufficient to demonstrate the fundamentals of our approach, resolving this limitation would substantially improve the usability of our system by allowing for both smaller QR codes and a higher level of error correction. Experimentally we found that smaller version number QR codes were much faster to read, and less error prone, than their higher capacity counterparts.

Regarding scalability, since our solution is decentralised and offline from any certificate issuer or revocation identity, there are inherently few issues to be considered. Nonetheless, each healthcare provider must maintain a database of user records which includes the $\mathit{CID}$ of their antibody certificate. In practice healthcare providers must already maintain patient records on the same scale as is required for antibody certificates and the addition of an extra field should not be problematic. For revocation, service providers must download the list of revoked $\mathit{CID}$s from each healthcare provider and users must download the list of revoked verifiers. This can easily be automated and is amenable to alternative techniques such as OCSP \cite{DBLP:journals/rfc/rfc6960} should the scale of revocation prove to contraindicate basic revocation lists in practice.

Lastly, while conducting usability studies of our approach is outside the scope of this work, we believe that the simple user interface shown in Fig. 2 is encouraging in this regard. We note that key management in SecureABC is, for users, a relatively straightforward matter of looking after their QR code in the same way as they might a photocard driver’s license, a bank card or indeed -- a mobile phone. Lost or stolen certificates are handled by contacting the healthcare issuer to receive a replacement.

\section{Ways to Mitigate Discrimination} \label{sec:non-discrimination}

The SecureABC system we present in Section \ref{sec:secureabc}, and indeed all forms of antibody certification, rely on policy and regulation to ensure that any resulting immunity-based discrimination is both proportional and necessary. In this section we are motivated to explore the ways in which individual discrimination could be technically mitigated, in a privacy preserving way, whilst still providing a mechanism for computing aggregate information about infection risks. Aggregate infection risk statistics are already being used by countries that have adopted ``local lockdown'' mechanisms which monitor regional COVID-19 risk and accordingly adjust policy. Currently, this information comes from local testing and reporting\footnote{\url{https://data.london.gov.uk/dataset/coronavirus--covid-19--cases}} and is therefore unable to capture transmission risks at the level of individual transport routes, workplaces or stores.

\vspace{1ex}
\noindent\textbf{Use case 4: aggregate transmission risk statistics.}
A national healthcare authority seeks to localise and mitigate COVID-19 outbreaks at the level of individual physical spaces. For this purpose, people entering a space will provide a health token that will allow the service provider to compute the aggregate sum of transmission risk over some arbitrary period such as a day or a week. Such statistics might help to reduce the need for more extensive regional or even national restrictions by allowing for more precisely targeted interventions. \vspace{1ex}




We first present two potential solutions for realising such health tokens before discussing the trade-offs with respect to discrimination. 

\subsection{Randomised Health Tokens}
\label{sec:rand_tokens}

Here we present the concept of a \textit{randomised health token} for transmission risk; a randomised health token that allows the collective transmission risk posed by groups of users to be estimated  whilst also protecting individuals from immunity-based discrimination.

To realise randomised health tokens we employ randomised response, a commonly used method to provide differential privacy, to issue users with health tokens based on their ``immunity status'' or transmission risk. Each user's transmission risk is potentially randomised but can nonetheless be used to reconstruct the underlying distribution of real risk over a sufficiently large number users. Before defining our randomised health token protocol, we first detail our randomised response protocol $\mathit{protocol}_{\mathit{DP}}$.

\subsubsection*{Randomised response protocol}

Let $\mathit{protocol}_{\mathit{DP}}$ take as input the user's true transmission risk status $i_{\mathit{true}} \in \{0, \dots, k-1\}$ which can be one of $k$ options and $\epsilon$. Here $\epsilon$ is a measure of the trade-off between privacy and accuracy; a low value of epsilon provides high privacy, since each user's risk status is more likely to be noise, but reduced utility for much the same reason. The $\mathit{protocol}_{\mathit{DP}}$ outputs $i_{\mathit{true}}$ with probability $\frac{e^{\epsilon}-1}{e^{\epsilon}+k-1}$, else we select a uniformly random response from $\{0, \dots, k-1\}$. Clearly it is possible to select a response at random and to still provide the true value $i_{\mathit{true}}$ by chance. 

An unbiased estimate $\hat{f}$ of the frequency $f$ of a given risk status $i_{\mathit{true}} \in \{0, \dots, k-1\}$ is calculated as follows:
\begin{equation}\label{eq:debias}
\hat{f}=\frac{e^{\epsilon}+k-1}{e^{\epsilon}-1}(\tilde{f}-1/k).
\end{equation}

\subsubsection*{Randomised health tokens protocol}

The randomised health tokens protocol comprises three phases: issue, authentication and aggregation.

\vspace{1ex}

\noindent \textbf{Issue}
\begin{enumerate}
	\item $H$ assesses $A$'s transmission risk indicator to be $i_{\mathit{true}}$. 
	\item $H$ computes $\mathit{i}_{\mathit{DP}} \leftarrow  protocol_{DP}(i_{\mathit{true}}, \epsilon)$ and composes $$\mathit{token}_{\mathit{DP}} = \mathit{sign_{PK_H}}\{\mathit{\mathit{i}_{\mathit{DP}}, date_{issue}, date_{end}}\}$$ where $\mathit{date_{issue}}$ and $\mathit{date_{end}}$ are the date the health token is issued and expires, respectively.
	\item $A$ receives her true rest result $i_\mathit{true}$ and her randomised health token $\mathit{token}_{\mathit{DP}}$.
\end{enumerate}

\noindent \textbf{Authentication}
\begin{enumerate}
	\item $A$ presents $\mathit{token}_{\text{DP}}$ to $V$ when accessing a physical space.
	\item $V$ verifies $H$'s signature on $\mathit{token}_{\text{DP}}$, validates $date_{end}$ and, if successful, records the randomised transmission risk $\mathit{i}_{\mathit{DP}}$. 
\end{enumerate}

\noindent \textbf{Aggregate}

\begin{enumerate}
	\item Periodically, $V$ counts the frequency $f$ of each value of $\mathit{i}_{\mathit{DP}}$ it has received during the period. $V$ then debiases the frequency estimates as per Equation \ref{eq:debias} to recover an unbiased transmission risk distribution.
\end{enumerate}

Each randomised health token is cryptographically signed by $H$ and therefore cannot be forged by the user. The bitstring representation of each signature will be unique (except for a negligible probability) for each health token and therefore constitutes a unique identifier. This unique identifier may be used to revoke health tokens that are misused, for example using a privacy-preserving technique such as \cite{DBLP:conf/ccs/NaorPR19}. Note that regardless of the randomised transmission risk value $i_\mathit{DP}$, $A$ still learns her true transmission risk $i_\mathit{true}$ during the issue phase of the protocol. This method allows service providers to obtain an accurate estimate of the risk they have encountered, while providing plausible deniability to all users whose tokens are authenticated. 



\subsection{Secret-Shared Health Tokens}
If one or more additional trusted parties can be assumed, then privacy-preserving transmission risk aggregates can also be computed without the requirement of adding noise to individual risk values. Our method is to use a Linear Secret Sharing Scheme (LSSS) to encrypt user transmission risk values. These encrypted values will enable the service provider $V$ and an additional third-party helper server $W$ to jointly compute the sum of many users transmission risk. Provided that at least one of these two parties remains honest then secret-shared health tokens provide stronger privacy than randomised health tokens -- the system leaks nothing about a user’s transmission risk to the service provider, except what can be inferred from the final sum.



\subsubsection{Secret-shared health tokens protocol} Let $p$ denote a large prime and $(pk_W, sk_W)\leftarrow \mathit{key\text{-}gen_{enc}}(\cdot)$ denote the public key of an independent third-party $W$ that will assist in the protocol. Furthermore, let both $V$ and $W$ initialise an accumulator value, $j_V = 0$ and $j_W = 0$ respectively, which will be used to sum the shares in each reporting period. For example, $W$ could be a university or a health organisation separate from the government. Then the secret-shared health tokens protocol comprises three phases: issue, authentication and aggregation which are defined as follows.


\vspace{1ex}

\noindent \textbf{Issue}
\begin{enumerate}
	\item $H$ assesses $A$'s transmission risk indicator to be $i_{\text{true}}$.
	
	\item $H$ computes two random shares $[i_{\text{true}}]_V$ and $[i_{\text{true}}]_W$ such that $i_{\text{true}} = [i_{\text{true}}]_V + [i_{\text{true}}]_W \mod p$. 
	
	\item $H$ encrypts $[i_{\text{true}}]_W$ using $W$'s public key $pk_W$ and then signs the resulting encrypted secret share: $$\mathit{share}_W = \mathit{sign_{PK_H}} \big \{ \mathit{enc}_{pk_W}([i_{\text{true}}]_W)\big \}$$
	
	\item Let $\mathit{share}_V = [i_{\text{true}}]_V$, then $H$ composes the secret-shared health token as:
	$$\mathit{token}_{\text{SS}} = \mathit{sign_{PK_H}} \{ \mathit{share}_W, \mathit{share}_V, date_{\text{issue}}, date_\text{end}\}$$
	
	
	

	\item $A$ receives her true rest result $i_\text{true}$ and her secret-shared health token $\mathit{token}_{\text{SS}}$.
\end{enumerate}

\noindent \textbf{Authentication}
\begin{enumerate}

	\item $A$ presents $\mathit{token}_{\text{SS}}$ to $V$ when accessing a physical space.
	
	\item $V$ verifies $H$'s signature on $\mathit{token}_{\text{SS}}$ and validates $date_{end}$. $V$ aborts this phase of the protocol if this step fails.
	
	\item $V$ sends sends $\mathit{share}_W$ to $W$ and then adds $\mathit{share}_V$ to its local accumulator value for this period: $j_{V} = j_{V} + \mathit{share}_V \mod p$
	
	\item $W$ receives $\mathit{share}_W$, verifies $H$'s signature, decrypts the secret-shared token using $sk_W$ and then adds the result to its local accumulator value for this period:  $j_{W} = j_{W} + [i_{\text{true}}]_W \mod p$.
	
\end{enumerate}

\noindent \textbf{Aggregate}

\begin{enumerate}
	\item $V$ and $W$ publish their accumulator values $j_{V}$ and $j_{W}$, respectively. Computing the sum of these values yields the sum of all of the transmission risk values provided by each user during the reporting period:
	
	$$j_{V} + j_{W} = \sum_i i_{\text{true}} \mod p$$
	
\end{enumerate}

As the end of the aggregation phase, both $V$ and $W$ learn the sum $\sum_i i_{\text{true}} \mod p$ but they learn nothing more about the user's private transmission risk values. Since all secret shares are signed by H, this protocol is robust against adversarial user inputs that could aim to corrupt the sum. 


\subsection{Utility-Privacy Trade-Off}

The two protocols described previously demonstrate that it is both technically feasible and efficient to retain some utility in systems which mitigate immunity-based discrimination. Here we are motivated to explore this trade-off between the utility of individual discrimination and the privacy of individual transmission risk. We compare the three protocols we specify in this work -- SecureABC, randomised health tokens, and secret-shared health tokens;  under four categories: (1) health discrimination, (2) level of accuracy, (3) the necessity of binding and (4) the need for an additional trusted entity beyond $H$. Table \ref{table:trade-off} summaries the trade-offs which we now consider individually in detail.

\begin{table}[h]
\begin{tabular}{|c|c|c|c|}
\hline
                                                                             & SecureABC & \begin{tabular}[c]{@{}c@{}}Randomised\\Tokens \end{tabular} & \begin{tabular}[c]{@{}c@{}}Secret-shared\\Tokens \end{tabular} \\ \hline
\begin{tabular}[c]{@{}c@{}}Mitigates\\Discrimination \end{tabular}
                                                          & \xmark        & \cmark       & \cmark  \\ \hline
\begin{tabular}[c]{@{}c@{}}Ideal\\Accuracy\end{tabular}                   & \cmark       & \xmark        & \cmark  \\ \hline
\begin{tabular}[c]{@{}c@{}}Binding\\Necessary\end{tabular}                  & \cmark       & \xmark        & \xmark   \\ \hline
\begin{tabular}[c]{@{}c@{}}Trusted\\Entities\end{tabular} & 1+        & 1+        & 2+  \\ \hline
\end{tabular} 
\vspace{2ex}
\caption{Utility-privacy trade-offs of the protocols we present in the paper.} 
\label{table:trade-off}
\vspace{-5ex}
\end{table}

\begin{description}[leftmargin=2ex, itemsep=1ex, topsep=2ex]
    \item[Mitigates Discrimination] All traditional antibody certificate systems, including the SecureABC protocol, are discriminatory by design. A user is strongly bound to their certificate and only good governance can ensure the necessity and proportionality of its usage. Whilst discrimination may of course be bad in many circumstances, as we discuss in Section \ref{sec:general_principles} there could be use cases where it is preferable to a worse alternative. In contrast, both our randomised and secret-shared health token proposals mitigate discrimination by concealing individual health records and only allowing aggregates of risk to be computed. The utility-privacy trade-off here is that these privacy-preserving techniques sacrifice their utility in cases where it is imperative to distinguish individuals.
    
    \item[Ideal Accuracy] Both SecureABC and secret-shared health tokens provide the ideal level of accuracy. In other words, the risk or antibody status assigned to users by their healthcare provider communicated without error to the service provider. In randomised health tokens, privacy is provided by adding noise (and hence deniability) to each users' transmission risk and therefore ideal accuracy is not realised. As we show in Figure \ref{fig:average_error2}, this error becomes increasingly small as the number of users over which aggregate risk is computed becomes large.
    
    \begin{figure}[h]
    \vspace{-0ex}
	\centering 
	\includegraphics[scale=0.60]{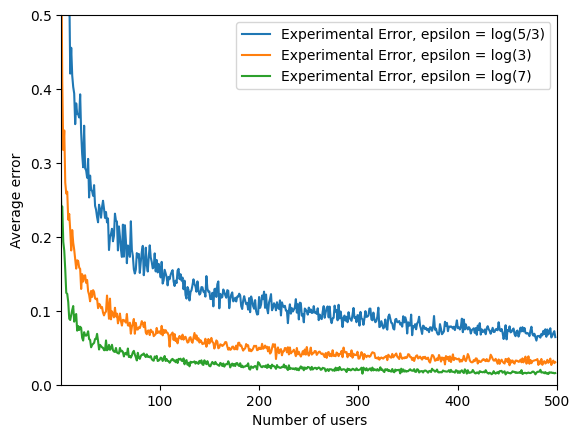}\vspace{-1ex}
	\caption{The average error introduced by our system for a given number of users. We let $k = 2$ and plot the error for $\epsilon = \; \mathrel{log}(\frac{5}{3}), \epsilon = \; \mathrel{log}(3), \epsilon = \; \mathrel{log}(7)$ and $k = 2$.}
	\label{fig:average_error2}
    \end{figure}
    
    \item[Binding Necessary] Binding to the owner of a certificate is necessary when the certificate entitles the bearer to special access or privilege. Since SecureABC allows antibody status to be proven by design, it also lends itself to entitling the bearer of a favourable certificate to special privileges. For example, a SecureABC credential could be required for access to jobs which involve working with vulnerable people. By comparison our randomised and secret-shared health tokens are about collective risk rather than individual privilege, and therefore may not necessarily need binding to the individual. The utility-privacy trade-off is that the any solution which requires binding must compromise privacy to the extent that an individual needs to strongly prove ownership. Whilst there are solutions that accomplish this without disclosing personal information such as a name or photograph, they ultimately still require some accountability to the individual.
    
    \item[Trusted Entities] All of our protocols require at least one root of trust. The healthcare authority $H$ must be trusted to issue antibody tests, collect the results and digitally sign them into credentials that are given to users. For app-users of the SecureABC protocol, $H$ also allows them to verify the service provider before showing their antibody certificate. In addition, the secret-shared health tokens protocol requires at least one additional third-party to be trusted to help aggregate secret shares in cooperation with each service provider. The trade-off of secret-shared health tokens is that this additional trusted entity is required to ensure discrimination on the basis of individual risk values is mitigated. 
    
\end{description}

These trade-offs between utility, privacy and individual discrimination motivate us to suggest that a combination of both SecureABC and either randomised or secret-shared health tokens could provide enhanced outcomes.  Integrating such a system would hinge on establishing a sensible public policy which specifies one of the two protocols on a case-by-case basis. An app could allow the user to provide either an antibody certificate or a health token depending on the minimum necessary access rights of each particular service provider. 

\section{Related Work} \label{sec:related_work}
Here we review the small number of alternative antibody certificate schemes which have been proposed. To the best of our knowledge, all alternative schemes are either based on a centralised architecture or propose the use of a blockchain. Many of the commercial systems being trialed and implemented by governments \cite{estonia_passport, uk_germany_passport} provide few technical details and cannot be fully understood or reviewed.

In the centralised category Estonia's antibody certificate system \cite{estonia_passport} enables people to share their so-called immunity status with a third-party using a temporary QR-code that is generated after authentication. Commercially, CoronaPass \cite{coronapass} also propose a centralised antibody certificate solution where service providers verify each user passport against a central database. Whilst security and legal measures can be put in place, in both these solutions, to deter the central authority from misusing the data they hold, it nonetheless represents an avoidable risk and a central point of failure. Involving the central party in each authentication risks large-scale user tracking and feature creep.

Eisenstadt et al. \cite{DBLP:journals/corr/abs-2004-07376} propose an antibody certificate scheme in which W3C-standard “verifiable credentials” \cite{sporny2019verifiable}, the “Solid” platform for decentralised social applications \cite{DBLP:conf/www/MansourSHZCGAB16} and a federated consortium blockchain are combined. In this system, a hash of each user’s certificate is stored in a consortium blockchain which is checked each time that an authentication between a user and verifier takes place. Many commercial antibody certificate solutions also indicate that a blockchain is included. The “Immupass Covid-19 Immunity Certificate” \cite{immupass} stores details of the tested individual directly in a consortium blockchain. User's present a QR code and their passport to a verifier and the corresponding test result and its validity is retrieved from the blockchain. CERTUS \cite{certus_blockchain_solution} uses a similar approach as does Vottun who are partnering with PwC for a trial in Spain \cite{lovelace_pwc}.

In the broader security and privacy literature relating to COVID-19, we note that there has been significant debate over the merits of centralised versus decentralised systems for digital contact tracing \cite{cryptoeprint:2020:531, europe_app, central_decentral_contact_trace2, DBLP:journals/corr/abs-2004-04059}. This is a complex debate because there is a tradeoff between privacy and utility. In particular, centralised contact tracing data may be a valuable tool in tackling the virus. We do not think this debate directly applies here because there is less utility in a centralized record of when and where antibody certificates have used. Consequently, we believe decentralisation is the best approach for antibody certificates.

In relation to the societal implications of antibody certificates, and our general principles which we present in Section \ref{sec:general_principles}, the Ada Lovelace Institute published a “Rapid evidence report” \cite{lovelace_report} which explores how non-clinical measures can be used to attempt to relax current governmental controls and restrictions without an intolerable rise in COVID-19 cases. Finally, we comment that, as we have discussed, the scientific evidence and public understanding of the situation is fast changing, therefore we concur with Eisenstadt et al. \cite{DBLP:journals/corr/abs-2004-07376} who call for regular reviews and oversight of sch processes to be handled by an Ethical Committee.

\section{Conclusion} \label{sec:conclusion}

In this work we explore the controversial technique of antibody certificates and present SecureABC: a decentralised, privacy-preserving system for issuing and verifying antibody certificates. This work is a necessary and important first step towards enabling an open discussion about the technical implications of antibody certificates. We consider the implications of antibody certificate systems, develop a set of general principles and security requirements for their deployment and show that these may be satisfied in practice.

Whilst the work here specifically considers antibody certificates for COVID-19, our framework of security properties, implementation results and privacy-preserving protocols contribute more generally to the development of any health certificate solution.

\begin{acks}

The authors are grateful to Marc Eisenstadt, Charles Raab, Markus Kuhn and Joseph Bonneau for their valuable feedback which has helped to improve this work.
\end{acks}

\bibliographystyle{plain}
\bibliography{main}

\end{document}